\documentclass{article}
\usepackage{tcolorbox}
\usepackage{amsthm}
\usepackage{hyperref}

\newcommand{\ReLU}{\mathrm{ReLU}}

\title{SQL4NN: Validation and expressive querying of models as
data}
\author{Mark Gerarts\thanks{Corresponding
author: mark.gerarts@student.uhasselt.be}
  \and Juno Steegmans\thanks{juno.steegmans@uhasselt.be}
  \and Jan Van den
  Bussche\thanks{jan.vandenbussche@uhasselt.be}}
\date{Data Science Institute, Hasselt University, Belgium}
\begin{document}
\maketitle
\begin{abstract}

We consider machine learning models, learned from data, to be an
  important, intensional, kind of data in themselves.  As such,
  various analysis tasks on models can be thought of as queries
  over this intensional data, often combined with extensional
  data such as data for training or validation.  We demonstrate
  that relational database systems and SQL can actually be well
  suited for many such tasks.

\end{abstract}
\section{Introduction}
Any serious machine learning project will quickly produce a multitude of
models learned from data.  These models are then validated,
tested in different ways, modified, retrained, and archived or deployed.
This multitude of models also constitutes valuable data in
itself.  We may refer to such data as \emph{intensional}, in
contrast to what we already know as \emph{extensional} data: training data,
background data, data for validation, etc.

Our point of view in this paper is that \emph{models are data
too} and should be managed using database technology, just like
the extensional data.  In particular, we should be able to
\emph{query} models.  Indeed, many tasks that we usually consider as 
validation, analysis, explanation, verification, pruning, etc.,
of models
\cite{rudin-stop-explaining,aws-book,LiuALSBK21,molnar-book,pruning101},
can actually be
considered as expressive database queries over extensional and
intensional data.  We give some examples:

(1)    Evaluating specified models in the database on specified
    inputs in the database, e.g., for validation or debugging.
    This also includes comparing different models on given
    inputs.  This example will probably not be surprising to the
    reader, but note here that the models are not fixed and part
    of the query, but instead are intensional data, also stored
    in the database.

(2)    Feature attribution: given a model and an input, compute
    for each feature its relevance to the output of the model on
    that input.

(3)    Verifying a model on possibly unseen inputs.
    This goes a level up.  For example, given a model for a speed
    control, we can ask if it would ever output a speed above 70
    mph, on any possible input.  It is not clear whether a
    standard query language would be able to express such
    queries.

(4)    Function geometry reconstruction. This
    generalizes the previous example and makes possible various
    other analyses.  The function represented by a feedforward
    neural network with ReLU activations is piecewise linear.  We
    can ask for all breakpoints of the function and the pieces
    between them.  For example, this allows us to compute
    definite integrals of the function represented by a model
    \cite{ql4nn}.

(5)    Pruning models. For example, given a neural network, find all
    hidden units that can be pruned, since they give a very weak
    signal on all inputs from some training data set
    \cite{pruning-nn}.  Note that this kind of querying is again
    different, since we are looking inside the models,
    opening up the black boxes, so to speak.

We refer to queries of the kind illustrated in 1 and 2 above as
\emph{evaluation} queries; those like 3 and 4 as
\emph{verification} queries; and those like 5 as \emph{white-box}
queries.

\section{SQL as a model-query language}

Our goal is to demonstrate that SQL is, in principle, well-suited
to support expressive querying of models.  SQL can cover relevant
queries from all three types of our taxonomy (evaluation,
verification, and white-box queries).  We will focus in our demo
on a basic but fundamental class of models: feedforward neural
networks with ReLU activations \cite{goodfellow-book}.

Neural networks, both the architecture as a graph of nodes and
edges, and the weights, can be naturally represented in
relations.  It is not surprising that SQL can express
\emph{evaluation} queries; see Section~\ref{secrel}.
Using SQL for \emph{white-box} queries has been much less considered.
Our contribution is to propose the idea of
model querying, and to demonstrate it in practice, ``out of
the box'', on small but real-world neural networks.
Moreover, model querying presents a nice
application of \emph{recursive} queries in SQL, as we will see.

That \emph{verification} queries are expressible in SQL is more
surprising.  Our demo provides the practical confirmation of a
recent theoretical result (by Grohe, Standke, and two of us
\cite{ql4nn}) to this effect.  Specifically, the theorem
considers first-order logic over the reals, with linear
arithmetical constraints, to query real functions.  This is the
standard logic for expressing properties to be verified on
neural networks \cite{aws-book}.  For fixed network depths,
it is shown that this logic
can indeed be simulated in SQL, working over relations storing a
neural network representing the function to be queried.


\section{Related Work} \label{secrel}

Expressive querying of models stored as intensional data is, to
the best of our knowledge, an original idea.  At the same time,
this idea fits more broadly in the important general theme of
integrating machine learning tasks in database systems; there has
been much research on this theme, as covered in surveys and
monographs \cite{boehmbook,declarmlsurvey,ss-m-ml-lifecycle-survey}.

Specifically on applications of SQL to machine learning, there
has been interesting work on using and extending SQL to not only
invoke \cite{madlib}, but also train machine learning models
\cite{jermaine_linalgebra,ml2sql,sql4ml}.  In essence, the
latter also involves evaluation-type queries over models.
Our novel focus here is on using SQL for the analysis of
already learned models, stored as intensional data in the
database.

We also note that the general idea of querying 
intensional data, other than machine
learning models, has repeatedly surfaced in database research,
e.g., constraints \cite{kkr_cql}, schemas \cite{schemasql2},
or even queries as data \cite{stoneb_queldatatype,metasql}.

\section{Demo overview}

\smallskip

\begin{tcolorbox}
  A video presenting our demo is available:
  \url{https://vimeo.com/1042270724}.  Everything shown in the
  video is implemented.  The database system running SQL is DuckDB, and
  supporting code is in Python with supporting libraries,
  including PyTorch.
\end{tcolorbox}

\subsection{Deploying stored models} \label{secdeploy}

\subsubsection*{Evaluating a network}

A neural network can be
straightforwardly stored in two tables. Table $\sf Node(id,bias)$
contains all units: input, hidden, as well as output, with
their bias (input units have no bias; we set a default value
there).  Table $\sf Edge(src,dst,weight)$ stores the
weights of the connections between units.  By adding a column
$\sf model\_id$ to both tables, we can also store multiple models.

An input vector can be stored in a table $\sf
Input(in\_id,val)$, giving values to the input units.  (Note that the
input units can be readily identified 
by querying for nodes without an incoming edge.)  Again, multiple
input vectors can be represented by adding a column $\sf vec\_id$.

When evaluating a network on an input vector, every unit $u$ from the
first hidden layer gets as value
\begin{equation}
  \label{evalunit}
  \sigma
(u.\mathsf{bias} + \sum_{(v,u,w) \in
  \mathsf{Edge}} w \cdot v.\mathsf{value}),
\end{equation}
with $\sigma$ an activation function expressible in SQL; we will
work with
$\ReLU(x)=\max(0,x)$.  Here, $v.\mathsf{value}$ is the value
given to input unit $v$ which can be obtained by joining with
the $\sf Input$ table.  This is straightforward to express in
SQL, resulting in a view $\sf Layer1(id,val)$.
Subsequent layers are evaluated similarly.

\subsubsection*{Demo: MNIST}

We have explained that, when the depth of the network is known,
evaluation can be expressed by a fixed composition of SQL views.
In our demo, we show this using the PyTorch-supplied example
network trained on the MNIST dataset for handwritten digit
recognition.  This is a convolutional neural network, but we
transform it to a graph structure to store it in relations.  Also,
the output layer does not use ReLU, but softmax, which is
expressible in SQL using arithmetic expressions and max
aggregation.  In our demo, we apply this to recognizing
handwritten zip codes.  We also show how to compare multiple
models (variants of the MNIST network of different sizes) on
digits drawn by hand by the user during the demo.  All this runs
interactively in SQL.

\subsubsection*{Recursion} 
\begin{figure}
  \newcommand{\krimp}{0.26}
  \newlength{\tussen}
  \setlength{\tussen}{1.5em}
  \newlength{\omlaag}
  \setlength{\omlaag}{2cm}
  \centering
  {\small (a)}
  \raisebox{-\omlaag}{\scalebox{\krimp}{\includegraphics{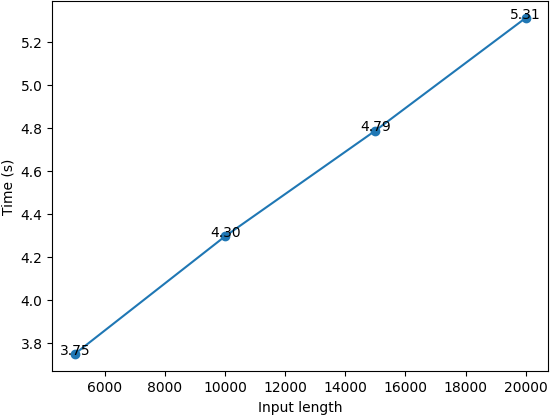}}}
    \hspace{\tussen}
  {\small (b)}
  \raisebox{-\omlaag}{\scalebox{\krimp}{\includegraphics{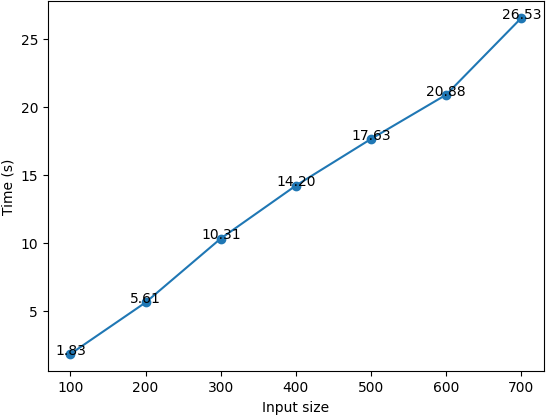}}}

  \bigskip

  {\small (c)}
  \raisebox{-\omlaag}{\scalebox{\krimp}{\includegraphics{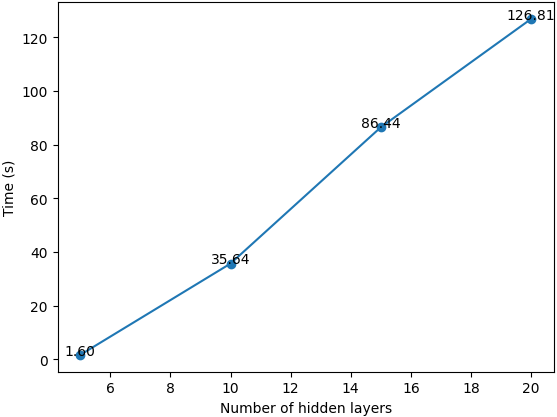}}}
    \hspace{\tussen}
  {\small (d)}
  \raisebox{-\omlaag}{\scalebox{\krimp}{\includegraphics{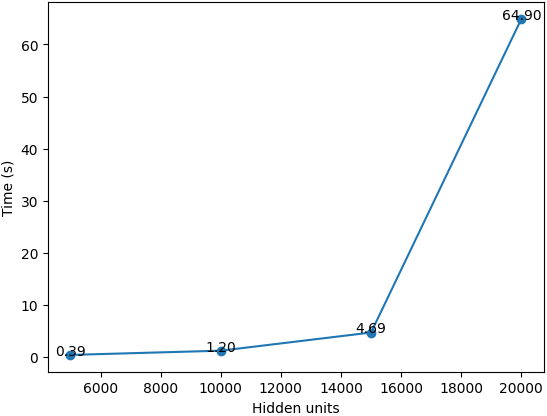}}}

  \bigskip

  {\small (e)}
  \raisebox{-\omlaag}{\scalebox{\krimp}{\includegraphics{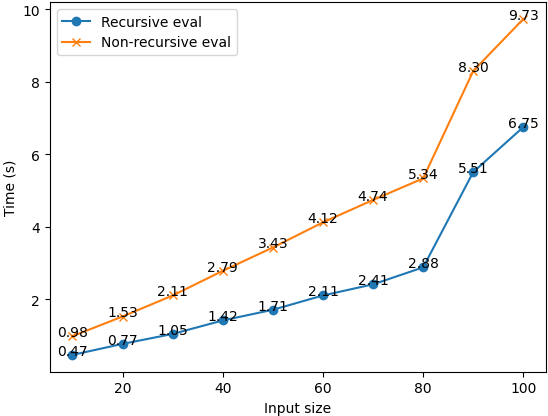}}}
\caption{(a)--(d): Scaling
input vector length; number of input vectors
to be evaluated; depth of the network;
number of hidden neurons in every layer.
(e) Evaluating a depth-5 network on multiple input
vectors is faster with a recursive SQL view
than with a fixed composition of 5 non-recursive views.}
  \label{perf}
\end{figure}

When the depth of the stored network
is not known, we cannot define a fixed number of views $\sf
Layer1$ to $\sf Layer5$, say.  Instead, a single view $\sf
Values(id,val)$ with values for all units in all layers must be
constructed progressively, layer by layer.  Thereto, we use SQL
recursion (recursive with-clause).  DuckDB supports recursive
views involving aggregation, which is crucial for our purposes.
Interestingly, the evaluation query is faster with recursion than
without (on a network of known depth), as shown in
Figure~\ref{perf}(e).

We further analyzed the running time of the recursive evaluation
query in an experimental setting similar to the MNIST example
($28\times 28$ input units, 4 layers, 10K hidden units in each
layer).  We then varied input length, input size, depth, and
layer size independently.  The results are shown in
Figure~\ref{perf}(a--d), showing linear scaling in all parameters
except for layer size, which indeed is quadratic: since we are
considering a fully connected feedforward network, there are
$s^2$ edges between successive layers, where $s$ is the layer
size.  Since our demo is meant to be interactive, both for this
experiment and the demo video we run everything on a commodity laptop (AMD
Ryzen 7 at 3.6GHz with 16GB RAM and 150GB SSD free space running Ubuntu
22.04.4).

\subsection{Model geometry reconstruction}

We demonstrate verification querying in SQL by showing how to
reconstruct the geometry of a function defined intensionally by a
stored ReLU network.  Even with a single hidden layer, such
networks can approximate any continuous function on a compact
interval \cite{pinkus-survey}.  Due to the use of ReLU, the
represented function is always piecewise linear.
So let us consider networks with a single input, one
hidden layer (arbitrarily large), and a single
output.  As before, the network is stored in relations $\sf Node$
and $\sf Edge$.

\begin{figure}
  \centering
  \scalebox{0.6}{\includegraphics{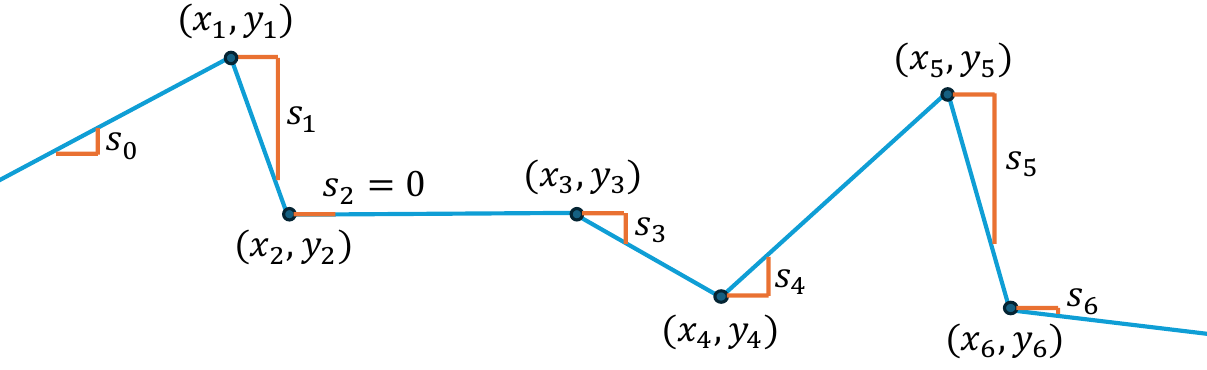}}
  \caption{Breakpoints and slopes. The blue line is the
  represented function.}
  \label{breakfig}
  \end{figure}

As the function represented by the network is piecewise linear,
we can reconstruct its geometry in SQL if we are able to define views
$\sf Breakpoints(x,y)$ and $\sf Slopes(x,s)$ containing the
breakpoints between the pieces, and the slopes of the pieces, as
illustrated in Figure~\ref{breakfig}.  (We also need the
initial slope $s_0$ which can be a separate singleton view.)
These views allow us to verify various
properties of the network. For example, querying the views for the initial
slope, the maximum $y$, and the slope of the last piece, we can
easily check if the function ever exceeds some threshold value
(compare the speed control example from the Introduction).

Expressing these views is an interesting application of SQL\@.
We can identify breakpoints with the hidden units, which are
readily retrieved from the $\sf Node$ and $\sf Edge$ columns.
Taking into account equation~\ref{evalunit} in
Section~\ref{secdeploy}, the $x$-coordinate of the
breakpoint generated by unit $u$ equals $-u.{\sf bias}/w$, where
$(v,u,w) \in {\sf Edge}$ and $v$ is the input unit.  The
$y$-coordinates can be computed using the evaluation query we
already discussed above.  Slopes can be determined by querying
for successive breakpoints and using the high-school math formula
for slope.  Thus, this is all well expressible in SQL.

\subsubsection*{Demo: sine wave, computing integrals}

We have stored in our database a neural network, trained
beforehand, that approximates a sine wave.  The demo first
visualizes the piecewise linear function represented by the
network, closely following the sine function.  We then 
retrieve the geometry (pieces) by interactively
materializing the SQL views $\sf Breakpoints$ and $\sf Slopes$.

To illustrate that these views allow a variety of further
verifications, we also demonstrate an SQL query that computes the
definite integral of the piecewise linear function.  The integral
of a sine wave over one period is 0, so the correct result
returned by SQL is readily verifiable by the demo audience.

\subsection{White-box querying}

White-box querying refers to querying and inspecting the model structure
directly.  For example, we can query for
the number of neurons, the number of distinct connections, or
the depth of the network (using recursion). But there is more.

\subsubsection*{Pruning}

Pruning machine learning models to make them smaller without losing
effectiveness is a huge topic \cite{pruning101}.  A basic but
effective form of pruning neural networks
consists of removing connections (edges) with weight
close to zero \cite{pruning-nn}; this may leave neurons
unconnected which may then be pruned.

These are easy queries on
the $\sf Edge$ relation.  For example, the following query
returns all nodes that have only incoming edges with weights
close to zero:
($\epsilon>0$ is a small parameter)
\begin{verbatim}
select dst from Edges group by dst
having -epsilon < min(weight) and max(weight) < epsilon
\end{verbatim}

\subsubsection*{Demo: interactive slider}
In the demo, we will consider a standard MNIST network 
and five smaller versions of it.  The original network has a few
prunable neurons, but going smaller decreases this number.  On
the other hand, increasing $\epsilon$ increases the number.
Using a slider for epsilon, the audience can play with these
effects.  The SQL queries run interactively.

\subsubsection*{Saliency maps}

\begin{figure}
  \centering
  \scalebox{0.5}{\includegraphics{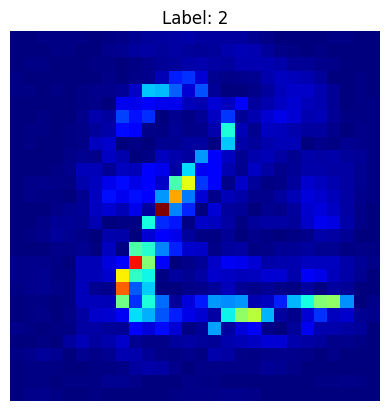}}
  \caption{Saliency map for a $28\times 28$ image of a handwritten digit.}
  \label{salfig}
  \end{figure}

Explaining the output of a model on a given input is a huge area
\cite{molnar-book,foscadino-explain-survey}.  
Here we demonstrate a very crude method, where we
compare the output with the output produced on a perturbation of the
input, with that component dropped or set to zero.  A measure of
the difference of the two outputs (e.g., the absolute value) may
be called the \emph{saliency} of the component under consideration
for that input.  When the input is an image, this results in a
``poor man's'' saliency map as illustrated in
Figure~\ref{salfig} (actual saliency map techniques
work with backpropagation).

The data for this Figure was indeed generated by SQL;
let us see how.
In Section~\ref{secdeploy} we have seen how to
define a view $\sf Values(id,val)$ returning the value ($\sf
val$) of each network unit ($\sf id$) when evaluating the network
on an input vector stored in relation $\sf Input(in\_id,val)$.  We can
expand this relation to contain a set of perturbed input vectors
as follows:
\begin{verbatim}
create view PInputs as
select I2.in_id as d_id, I1.in_id,
    (case when I1.in_id=I2.d_id then 0 else I1.val) as val
from Input I1, Input I2
\end{verbatim}
Here, $\sf d\_id$ is the ``dropped'' input unit, set to
zero.  This $\sf d\_id$ column now effectively
can play the role of a $\sf vec\_id$
(cf.~Section~\ref{secdeploy})
to store a set of input vectors in a
single relation.  We can easily generalize the SQL queries
for evaluating a network to evaluate such a set
of input vectors in parallel.  It now remains to join that result
with the result of evaluating the original input vector, and
to return the differences in outputs.
The exact SQL expressions will be visible during the demo.

\theoremstyle{remark}
\newtheorem*{remark}{Remark}

\begin{remark}
  Instead of ``dropping'' an input by setting it to zero, we may
  also really omit it from the neural network.
  This can be done by an alternative SQL approach where,
instead of modifying the $\sf Input$ table
  as done above, we modify the $\sf Edge$ table.  Specifically,
  we will define a view $\sf MEdges(d\_id,src,dst,weight)$ such
  that $\sf (src,dst,weight)$ is a tuple in the $\sf Edge$ graph
  where node $\sf d\_id$ is dropped.  Now $\sf d\_id$ can play the
  role of a $\sf model\_id$ (cf.~Section~\ref{secdeploy}) 
  so that $\sf MEdge$ stores a set of models.  As before we now
  evaluate all these models in parallel, joining with the original 
  $\sf Input$ relation.  Interestingly, the same approach also works
  for dropping hidden units, allowing to compute saliency or
  relevance of these as well.  Such queries qualify as highly
  expressive, true white-box queries.
\end{remark}

\subsubsection*{Demo: Saliency} In our demo we continue with the
MNIST model and let the audience draw a digit, upon which we
compute its classification in SQL as before, but also compute a
saliency map. The full and
self-contained SQL saliency query is rather complex
and takes too long to run interactively on a commodity laptop.
We can circumvent this by running the $28 \times 28$
perturbed evaluations separately, still each in SQL, but in a
concurrent loop governed by Python.

\section{Conclusion}

Much research is being done towards
in-database machine learning
(cf.~Section~\ref{secrel}).
We have shown that data systems supporting SQL also
support the further analysis of learned models (which often
integrates with learning, e.g., in the pruning application).  Our
current demonstration is a proof of concept and analyzes small
(but realistic) models running on a laptop; further
experimentation with state-of-the-art data systems for ML should
explore the handling of large models.
Other questions for further research are: Can graph
databases bring a benefit for our purposes, compared to
SQL\@?  How about other models, like decision trees
\cite{arenas-foil,arenas-dtfoil}, or other neural network
architectures?

\newcommand{\etalchar}[1]{$^{#1}$}

\end{document}